\documentclass[aps,pra,showpacs,superscriptaddress,preprint]{revtex4}

\usepackage{graphicx}
\usepackage{color}
\usepackage{lscape}
\usepackage{hyperref}
\usepackage{epstopdf}
\usepackage{lscape}
\usepackage{amsmath}
\usepackage{hyperref}
\usepackage{amsmath}
\usepackage{setspace}
\begin{document}

\title{Energy states of some diatomic molecules: Exact quantization rule approach}
\author{Babatunde. J. Falaye}
\email[E-mail: ]{fbjames11@physicist.net; babatunde.falaye@fulafia.edu.ng}
\affiliation{Applied Theoretical Physics Division, Department of Physics, Federal University Lafia,  P. M. B. 146, Lafia, Nigeria}
\author{Sameer M. Ikhdair}
\email[E-mail: ]{sikhdair@gmail.com; sameer.ikhdair@najah.edu}
\affiliation{Department of Physics, Faculty of Science, an-Najah National University,P. O. Box 7, Nablus, West Bank, Palestine }
\author{Majid Hamzavi}
\email[E-mail: ]{majid.hamzavi@gmail.com}
\affiliation{Department of Physics, University of Zanjan, Zanjan, Iran }
\date{%
%TCIMACRO{\TeXButton{today}{\today}}%
%BeginExpansion
\today%
%EndExpansion
}

\noindent
\begin{abstract}
\noindent
In this study, we obtain the approximate analytical solutions of the radial Schr\"{o}dinger equation for the Deng-Fan diatomic molecular potential by using exact quantization rule approach. The wave functions have been expressed by hypergeometric functions via the functional analysis approach. An extension to rotational-vibrational energy eigenvalues of some diatomic molecules are also presented. It is shown that the calculated energy levels are in good agreement with the ones obtained previously $E_{n\ell}-D$ (shifted Deng-Fan).

\noindent
{\bf Keywords}: Schr\"{o}dinger equation; exact quantization rule;  functional analysis; Deng-Fan potential.
\end{abstract}

\pacs{ 03.65.Ge; 03.65.-w; 02.30.Gp}

\maketitle

\section{Introduction}
According to the Schr\"{o}dinger formulation of quantum mechanics, a total eigenfunction provides implicitly all relevant information about the behaviour of a physical system. Thus, if it is exactly solvable for a given potential, an obtained eigenfunction can be used to describe such a system completely. This has made the exact solutions of quantum systems be an important subject and also attract much attention in the development of quantum mechanics [1-25].

Up to now, there have been several efficient methodology developed to find the exact solutions of quantum systems within the framework of the non-relativistic
and relativistic quantum mechanical wave equations. Few of these methods include the Feynman integral formalism \cite{I5,I16}, asymptotic iteration method (AIM) [1-5,17-21], functional analysis approach \cite{I22,I23,I24}, exact quantization rule method \cite{I25, I26, I27, I28, I29, N1, N2, N3,N4}, proper quantization rule \cite{I27,I30}, Nikiforov-Uvarov (NU) method \cite{I31, I32, I33, I34}, supersymetric quantum mechanics \cite{I36,I36,I37,I38,I39,I40,I41}, etc.

Very recently, Ma and Xu have proposed an exact (improved) quantization rule and applied it to calculate the energy levels of some exactly solvable quantum mechanical problems. These include the finite square well, the Morse, the symmetric and the symmetric and asymmetric Rosen-Morse, the harmonic oscillator, the first and second P\"{o}schl-Teller potentials \cite{I25}. By using this same rule, Dong and co-researchers \cite{I41} have obtained the analytical solutions of the Schr\"{o}dinger equation for the deformed harmonic oscillator in one dimension, the Kratzer potential and pseudoharmonic oscillator in three dimensions. The energy levels of all the bound states are easily calculated from this quantization rule.

\begin{figure}[!htb]
\centering \includegraphics[height=100mm,width=160mm]{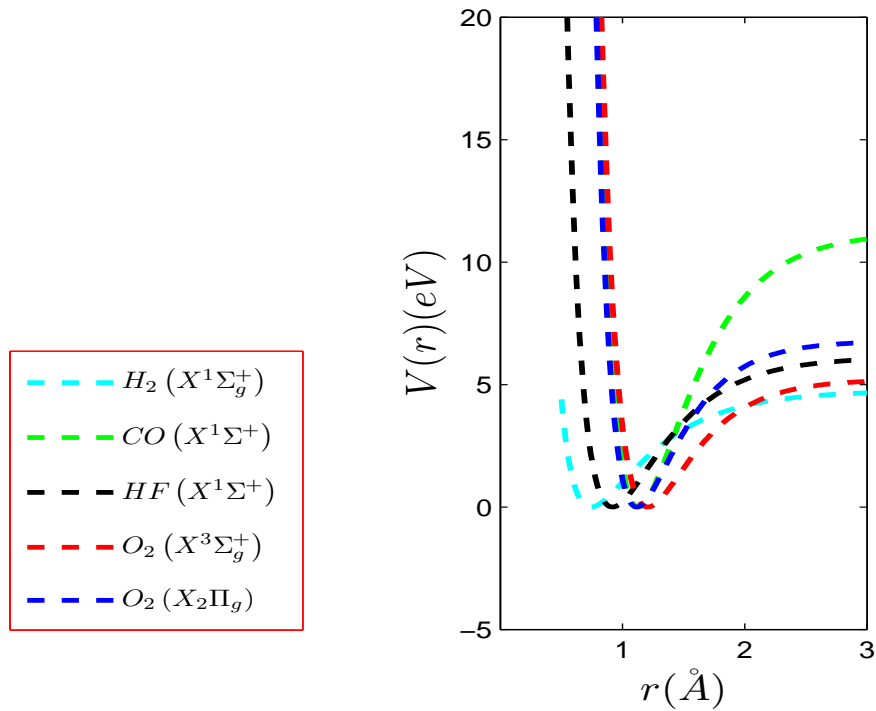}
\caption{{\protect\footnotesize (color : online) Shape of Deng-Fan diatomic molecular potential for different diatomic molecules.}}
\label{fig1}
\end{figure}

The priority purpose of the present work is to obtain the obtain the bound state solution of Deng-Fan diatomic potential model via this quantization rule. The Deng-Fan diatomic potential model also known as the generalized Morse potential which was proposed some decades ago by Deng and Fan \cite{I42}. This is done in an attempt to finding a more suitable diatomic potential to describe the vibrational spectrum, qualitatively similar to the Morse potential with the correct asymptotic behaviour as the inter-nuclear distance approaches zero. This potential model can be used to describe the motion of the nucleons in the mean field produced by the interactions between nuclei \cite{I43}. This potential has been used to describe diatomic molecular energy spectra and electromagnetic transitions and it is an ideal inter-nuclear potential in diatomic molecules with the same behavior for $ r \rightarrow 0 $. Because of its importance in chemical physics, molecular spectroscopy, molecular physics and related areas, the bound state solutions of the relativistic and non-relativistic wave equations have been studied by several authors \cite{I7,I29,I42,I43}. The shape of this potential with respect to some diatomic molecules is shown in figure{\ref{fig1}}.

The organization of this work is as follows. In the next section, we briefly introduced the exact quantization rule. In Section 3, we apply this quantization rule to obtain the bound state solutions of the Deng-Fan molecular potential. Section 4 present the numerical results and finally and a brief conclusion.
\section{Exact Quantization Rule}
In this section, the brief review of this quantization rule is given. The details can be found in Refs \cite{I25,I26}. It is well known that in one dimension, Schr\"{o}dinger equation 
\begin{equation}
\frac{d^2}{dx^2}\psi(x)+\frac{2\mu}{\hbar^2}\left[E-V(x)\right]\psi(x)=0,
\label{E1}
\end{equation}
can be written in the following form
\begin{equation}
\phi'(r)+\phi(x)^2+k(x)^2=0,\ \ \ \mbox{with}\ \ \ \ \ \ k(x)=\sqrt{\frac{2\mu}{\hbar^2}[E-V(x)]},
\label{E2}
\end{equation}
where $\phi(x)=\psi'(x)/\psi(x)$ is the logarithmic derivative of the wave function $\psi(x)$, the prime denotes the derivative with respect to the variable x, $\mu$ denotes the reduced mass of the two interacting particles, $k(x)$ is the momentum and $V(x)$ is a piecewise continuous real potential function of $x$. For the Schr\"{o}dinger equation, the phase angle is the logarithmic derivative $\phi(x)$. From equation (\ref{E2}), as $x$ increases across a node of wave function $\psi(x)$, $\phi(x)$ decreases to $-\infty$, jumps to $+\infty$ and then decreases again.

In the recent years, Ma and Xu \cite{I25, I26} generalized this exact quantization rule to the 3D radial Schr\"{o}dinger equation
with spherically symmetric potential by simply making the replacements $x\rightarrow r$ and $V(x)\rightarrow V_{eff}(r)$:
\begin{equation}
\int_{r_a}^{r_b}k(r)dr=N\pi+\int_{r_a}^{r_b}\phi(r)\left[\frac{dk(r)}{dr}\right]\left[\frac{d\phi(r)}{dr}\right]^{-1},\ \ \ \ k(r)=\sqrt{\frac{2\mu}{\hbar^2}[E-V_{eff}(r)]} .
\label{E3}
\end{equation}
where $r_a$ and $r_b$ are two turning points determined by $E=V_{eff}(r)$. The $N=n+1$ is the number of the nodes of $\phi(r)$ in the region $E_{n\ell}=V_{eff}(r)$ and is larger by one than the number $n$ of the nodes of wave function $\psi(r)$.  The first term $N\pi$ is the contribution from the nodes of the logarithmic derivative of wave function, and the
second is called the quantum correction.  It is found that for all well-known exactly solvable quantum systems, this quantum correction is independent of the number of nodes of wave function. This means that it is enough to consider the ground state in calculating the quantum correction, i.e.
\begin{equation}
Q_c=\int_{r_A}^{r_B}k_0'(r)\frac{\phi_0}{\phi_0'}dr
\label{E4}
\end{equation}
In the recent years, this quantization rule has been used in many physical systems to obtain the  exact solutions of many exactly solvable quantum systems \cite{I25,I26,I27,I28,I29,I30}
%%%%%%%%%%%%%%%%%%%%%%%%%%%%%%%%%%%%%%%%%%%%%%%%%%%%%%%%%%%%%%%%%%%%%%%%%%%%%%%%%
\section{Application to Deng-Fan molecular potential}
In this section, we apply the exact quantization to study the ro-vibrational energy states of some diatomic molecules. To begin, we write Schrodinger equation with The Deng-Fan diatomic molecular potential as
\begin{equation}
\frac{d^2R_{n\ell}(r)}{dr^2}+\frac{2\mu}{\hbar^2}\left[E_{n\ell}-D\left(1-\frac{b}{e^{ar}-1}\right)^2-\frac{\ell(\ell+1)\hbar^2}{2\mu r^2}\right]R_{n\ell}(r)=0,
\label{E5}
\end{equation}
where $n$ and $\ell$ denote the radial and orbital angular momentum quantum numbers, $r$ is the internuclear separation of the diatomic molecules, and $E_{n\ell}$ is the bound-state energy eigenvalues. The $\mu$ and $V(r)$ represent the reduced mass and interaction potential respectively. It is well-known that the equation of form (\ref{E5}) is an exactly solvable problem for s-wave. But for $\ell-$wave states, the problem is not analytically solvable. To obtain the bound state solutions, we must therefore resort to using an approximation similar to other works \cite{I44,I45,I46,I47} to deal with the centrifugal term or alternatively, to solve numerically. It is noted that, for short potential range, the following formula is a good approximation to the centrifugal term:
\begin{equation}
\frac{1}{r^2}\approx \alpha^2\left[d_0+\frac{e^{\alpha r}}{\left(e^{\alpha r}-1\right)^2}\right],
\label{E6}
\end{equation}
where $d_{o} = \frac{1}{12}$. It should be noted that  this approximation reduce to the one used by Dong and Gu \cite{I29}, Dong et al \cite{I23}, and Qiang and Dong \cite{I24} when $d_{o} = 0$. Now if we consider the approximation (\ref{E6}) and equation (\ref{E5}), we can find the effective potential as
\begin{equation}
V_{eff}(\varrho)=D+\frac{\ell(\ell+1)\alpha^2d_0\hbar^2}{2\mu}+\left(\frac{\ell(\ell+1)\alpha^2\hbar^2}{2\mu}-2Db \right)\varrho+\left(\frac{\ell(\ell+1)\alpha^2d_0\hbar^2}{2\mu}+Db^2\right)\varrho^2,
\label{E7}
\end{equation}
where we have introduced a new transformation of the form $\varrho=\frac{e^{-\alpha r}}{1-e^{-\alpha r}}$. Now, let us begin the application of the quantization rule to study the potential $V_{eff}(\varrho)$. To perform this task, we have to first calculate the turning points $\varrho_a$ and $\varrho_b$ determined by solving $V(\varrho) = E_{n\ell}$. Thus we have
\begin{eqnarray}
&&\varrho_a=-\frac{Q}{2R}-\frac{\sqrt{Q^2-4R(P-E_{n\ell})}}{2R},\ \ \varrho_b=-\frac{Q}{2R}+\frac{\sqrt{Q^2-4R(P-E_{n\ell})}}{2R}\nonumber\\
&&\mbox{with}\ \  k(\varrho)=\frac{\sqrt{2\mu R}}{\hbar}[(\varrho_a-\varrho)(\varrho-\varrho_b)]^{1/2}
\label{E8}
\end{eqnarray}
where $k(\varrho)$ is the momentum between the two turning points $\varrho_a$ and $\varrho_b$. We have also introduced three parameters $P$, $Q$ and $R$ for mathematical simplicity. These parameters are as follows:
\begin{equation}
P=D+\frac{\ell(\ell+1)\alpha^2d_0\hbar^2}{2\mu}, \ \ Q=\frac{\ell(\ell+1)\alpha^2\hbar^2}{2\mu}-2Db\ \ \mbox{and}\ \ \ R=\frac{\ell(\ell+1)\alpha^2d_0\hbar^2}{2\mu}+Db^2.
\label{E9}
\end{equation}
The non linear Riccati equation for the ground state is written in terms of the new variable $\varrho$ as
\begin{equation}
-a\varrho(1+\varrho)\phi_0'(\varrho)+\phi_0^2(\varrho)+\frac{2\mu}{\hbar^2}\left[E_{0\ell}-V_{eff}(\varrho)\right]\phi_0(\varrho)=0.
\label{E10}
\end{equation}
Thus, since the logarithmic derivative $\phi_0(\varrho)$ for the ground state has one zero and no pole, therefore we assume the following solution for the ground states
\begin{equation}
\phi_0(\rho)=\mathcal{A}+\mathcal{B}\varrho.
\label{E11}
\end{equation}
On substituting equation (\ref{E11}) into (\ref{E10}) and then solve the non-linear Riccati equation, we obtain the ground state energy as
\begin{equation}
E_{0\ell}=P-\frac{\hbar^2A^2}{2\mu}.
\label{E12}
\end{equation}
Also, $\mathcal{A}$ and $\mathcal{B}$ are as follows:
\begin{equation}
\mathcal{A}=\frac{\mu}{\hbar^2}\frac{Q-R}{B}+\frac{B}{2} \ \ \mbox{and}\ \ \ \mathcal{B}=\frac{a}{2}+\frac{1}{2}\sqrt{a^2+\frac{8\mu R}{\hbar^2}}. 
\label{E13}
\end{equation}
It is worth to be noted that we only choose the positive sign in front of the square root for $\mathcal{B}$. This is because the logarithmic derivative $\phi_0(\varrho)$ will decrease
exponentially, which is required physically. Let us now calculate the quantum correction. For this purpose, we utilize the integrals given by appendix A, and we obtain
\begin{eqnarray}
\int_{r_a}^{r_b}k_0'(r)\frac{\phi_0(r)}{\phi_0'(r)}dr&=&-\int_{\varrho_a}^{\varrho_b}\frac{k_0'(\varrho)}{\alpha\varrho(1+\varrho)}\frac{\phi_0(\varrho)}{\phi_0'(\varrho)}d\varrho=\frac{1}{\alpha}\sqrt{\frac{2\mu R}{\hbar^2}}\int_{\varrho_a}^{\varrho_b}\frac{[\mathcal{A}+\mathcal{B}]\varrho\left[\varrho-\left(\frac{\varrho_a+\varrho_b}{2}\right)\right]d\varrho}{\mathcal{B}\varrho(1+\varrho)\sqrt{(\varrho_a-\varrho)(\varrho-\varrho_ b)}}\nonumber\\
&=&\frac{1}{\alpha}\sqrt{\frac{2\mu R}{\hbar^2}}\int_{\varrho_a}^{\varrho_b}\frac{d\varrho}{\sqrt{(\varrho_a-\varrho)(\varrho-\varrho_ b)}}\left[\frac{\left(\frac{\mathcal{A}}{\mathcal{B}}-1\right)\left(\frac{\varrho_a+\varrho_b}{2}+1\right)}{1+\varrho}-\frac{\frac{\mathcal{A}}{\mathcal{B}}\left(\frac{\varrho_a+\varrho_b}{2}\right)}{\varrho}+1\right]\nonumber\\
&=&\frac{\pi}{\alpha}\sqrt{\frac{2\mu R}{\hbar^2}}\left[1+\frac{\sqrt{2\mu R}}{\mathcal{B}\hbar}\right].\label{E14} 
\end{eqnarray}
Furthermore, the integral of the momentum $k(r)$ can be found as follows:
\begin{eqnarray}
\int_{r_a}^{r_b}k(r)dr&=&-\int_{\varrho_a}^{\varrho_b}\frac{k(\varrho)}{\alpha\varrho(1+\varrho)}d\varrho=-\frac{1}{\alpha}\sqrt{\frac{2\mu R}{\hbar^2}}\int_{\varrho_a}^{\varrho_b}\frac{\sqrt{(\varrho_a-\varrho)(\varrho-\varrho_ b)}}{\varrho(1+\varrho)}\nonumber\\
&=&-\frac{\pi}{\alpha}\sqrt{\frac{2\mu R}{\hbar^2}}\left[\sqrt{(\varrho_a+1)(\varrho_b+1)}-1-\sqrt{\varrho_a\varrho_b}\right]\label{E15}\\
&=&-\frac{\pi}{\alpha}\left[\sqrt{\frac{R-Q+P-E_{n\ell}}{R}}-1-\sqrt{\frac{P-E_{n\ell}}{R}}\right],\nonumber
\end{eqnarray}
where we have used an appropriate standard integral in the appendix A. Now by combining the results obtained by equations (\ref{E14}) and (\ref{E15}) with equation equation (\ref{E3}), i.e.
\begin{equation}
-\frac{\pi}{\alpha}\left[\sqrt{\frac{R-Q+P-E_{n\ell}}{R}}-1-\sqrt{\frac{P-E_{n\ell}}{R}}\right]=N\pi+\frac{\pi}{\alpha}\sqrt{\frac{2\mu R}{\hbar^2}}\left[1+\frac{\sqrt{2\mu R}}{\mathcal{B}\hbar}\right],
\label{E16}
\end{equation}
the energy eigenvalues spectrum can then be found as
\begin{equation}
E_{n\ell}=D(b+1)^2+\frac{\ell(\ell+1)a^2\hbar^2d_0}{2\mu}-\frac{\hbar^2a^2}{2\mu}\left[\frac{\left(\mathcal{B} +\alpha n\right)}{2\alpha}+\frac{2\mu Db(b+2)}{2\hbar^2\alpha\left(\mathcal{B}+\alpha n\right)}\right].
\label{E17}
\end{equation}
Let us now obtain the corresponding wave function for this system. For this purpose, we introduced a new transformation of the form $z=e^{-\alpha r}$ $\in(e^\alpha, 0)$ in equation (\ref{E5}) which maintained the finiteness of the transformed wave functions on the boundary conditions we have
\begin{eqnarray}
&&z^2\frac{d^2R_{n\ell}(z)}{dz^2}+z\frac{dR_{n\ell}(z)}{dz}+\left(\frac{U+Vz+Wz^2}{(1-z)^2}\right)R_{n\ell}(z)=0 \nonumber\\ 
&&\mbox{with}\ W =\frac{2\mu}{\alpha^2\hbar^2}\left(E_{n\ell}-D\right)-\ell(\ell+1)d_0,\label{EE18}\\
&&V=\frac{4\mu bD}{\alpha^2\hbar^2}-\frac{4\mu}{\alpha^2\hbar^2}\left(E_{n\ell}-D\right)-\ell(\ell+1)(1+2d_0) \nonumber\\
&&\mbox{and}\ U=\frac{2\mu}{\hbar^2}\left(E_{n\ell}-D\right)-\frac{2\mu Db}{\alpha^2\hbar^2}(b+2)+\ell(\ell+1)d_0\nonumber.
\end{eqnarray}
Furthermore, equation (\ref{EE18}) is transformed into a more convenient second-order homogeneous linear differential equation via the following transformation
\begin{equation}
R_{n\ell}(z)=z^p(1-z)^{\frac{\mathcal{B}}{\alpha}}G_{n\ell}(z), \ \ \ \mbox{with}\ \ \ p=\frac{1}{2}+\sqrt{-W}.
\label{E19}
\end{equation}
Substituting of equation (\ref{E19}) into equation (\ref{EE18}), we can find
\begin{equation}
G_{n\ell}''(z)+G_{n\ell}'(z)\left[\frac{(2p+1)-z\left(2p+\frac{\mathcal{B}}{\alpha}+1\right)}{z(1-z)}\right]+G_{n\ell}(z)\left[\frac{\left(\frac{\mathcal{B}}{\alpha}+p\right)^2+U}{z(1-z)}\right]=0.
\label{E20}
\end{equation}
The solution to the above second order differential equation can be expressed in terms of  the hypergeometric function as:
\begin{equation}
G_{n\ell}(z)=\ _2F_1(\zeta,\eta;\gamma; z) \ \ \mbox{or}\ \ R_{n\ell}(z)=z^p(1-z)^{\frac{\mathcal{B}}{\alpha}}\ _2F_1(\zeta,\eta;\gamma; z).
\label{E21}
\end{equation}
The following notations:
\begin{equation}
\zeta=p+\frac{\mathcal{B}}{\alpha}-\sqrt{-U},\ \ \eta=p+\frac{\mathcal{B}}{\alpha}+\sqrt{-U}\ \ \mbox\ \ \gamma=2p+1
\label{E22}
\end{equation}
has been introduced in equation (\ref{E21}) so as to avoid mathematical complexity. By considering the finiteness of the solutions, $G_{n\ell}(z)$ approaches infinity unless $\zeta$ is a negative integer. This is an indication that $G_{n\ell}(z)$ will not be finite everywhere unless we take $p+\frac{\mathcal{B}}{\alpha}-\sqrt{-U}=-n$. Thus, $\eta$ given by equation (\ref{E22}) can be re-written in terms of this condition, and finally we can write the wave function as
\begin{equation}
R_{n\ell}(z)=N_{n\ell}z^p(1-z)^{\frac{\mathcal{B}}{\alpha}}G_{n\ell}(z)=z^p(1-z)^{\frac{\mathcal{B}}{\alpha}}\ _2F_1(-n, 2\left(p+\frac{\mathcal{B}}{\alpha}\right); 2p+1; z).
\label{E23}
\end{equation}
$N_{n\ell}$ is the normalization constant.
\section{Results and Conclusion}
By using the known spectroscopic values in Table \ref{tab1} we obtain the energy states of few selected diatomic molecules for various vibrational $n$ and rotational $\ell$ angular momentum as shown in Tables \ref{tab2} and \ref{tab3}. These  diatomic molecules are $H_2 \left(X^1\Sigma^+_g\right)$, $CO \left(X^1\Sigma^+\right)$,	$HF \left(X^1\Sigma^+\right)$, $O_2 \left(X^3 \Sigma^+_g\right)$ and $O_2 ^+\left(X^2 \Pi_g\right)$. The spectroscopic parameters are taken from the work of Kunc-Gordillo-Vazquez \cite{I48} and Oyewumi et al. \cite{I47}. We also applied following conversion $\mu/10^{-23}g = \mu\times6.0221415\times931.494028e6/c^2$ with $\hbar c = 1973.29eV\AA$ throughout our numerical computation.

In order to test the accuracy of our results,we give a numerical comparison of our obtained energy spectrum with the ones $H_2 \left(X^1\Sigma^+_g\right)$ and $CO \left(X^1\Sigma^+\right)$ obtained previously in the literature, for the shifted Deng-Fan potential (i.e. Deng-Fan potential shifted by the dissociation energy $D$). As it can been seen from Table \ref{tab2} our approximate results are in excellent agreement with the ones obtained previously by other authors and via other approach.  We also proceed furthermore to obtain the ro-vibrational energy spectrum (in $eV$ units) for the 5 selected diatomic molecules.

It is worth to be noted that the advantage of the method presented in this study is that, for exactly solvable quantum system, it enable one to find the energy spectrum directly in a simple way. Finally, we recommend that the extension of this method to obtain the bound state solution some other potentials such as: Wood-Saxon potential, Yukawa potential, Hellmann potential, etc.
\section*{Appendix A: Some useful standard integrals}
\begin{equation}
\int_{r_A}^{r_B}\frac{1}{\sqrt{(r-r_A)(r_B-r)}}dr=\pi \tag{A1}
\label{A1}
\end{equation}
\begin{equation}
\int_{r_A}^{r_B}\frac{1}{(a+br)\sqrt{(r-r_A)(r_B-r)}}dr=\frac{\pi}{\sqrt{(a+br_B)(a+br_A)}} .\tag{A2}
\label{A2}
\end{equation}
\begin{equation}
\int_{r_A}^{r_B}\frac{1}{r}{\sqrt{(r-r_A)(r_B-r)}}dr=\frac{\pi}{2}(r_A+r_B)-\pi\sqrt{r_Ar_B} .\tag{A3}
\label{A3}
\end{equation}
\begin{table}[!t]
{\scriptsize
\caption{Model parameters of the diatomic molecules studied in the present work. } \vspace*{10pt}{
\begin{tabular}{ccccc}\hline\hline
{}&{}&{}&{}&{}\\[-1.0ex]
Molecules(states)&$\mu/ 10^{-23}(g)$&$r_e (\AA)$&$D(cm^{-1})$& $a (\AA)$\\[2.5ex]\hline\hline
$H_2 \left(X^1\Sigma^+_g\right)$&0.084	&0.741	&38318&1.9506\\[1ex]
$CO \left(X^1\Sigma^+\right)$	  &1.146	&1.128	&90531&2.2994	\\[1ex]
$HF \left(X^1\Sigma^+\right)$	  &0.160	&0.917	&49382&2.2266\\[1ex]
$O_2 \left(X^3 \Sigma^+_g\right)$&1.337	&1.207	&42041&2.6636	\\[1ex]
$O_2 ^+\left(X^2 \Pi_g\right)$	&1.337	&1.116	&54688&2.8151	\\[1ex]
\hline\hline
\end{tabular}\label{tab1}}
\vspace*{-1pt}}
\end{table}
\begin{table}[!t]
{\scriptsize
\caption{ \small Comparison of the bound-state energy eigenvalues $-(E_{n\ell}-D)(eV)$ of $H_2$ and $C0$ molecules for various $n$ and rotational $\ell$ quantum numbers in Deng-Fan diatomic molecular potential. } \vspace*{10pt}{
\begin{tabular}{cccccccc}\hline\hline
{}&{}&{}&{}&{}&{}&{}&{}\\[-1.0ex]
$n$	&$\ell$	&Present ($H_2$)	&AIM \cite{I47}($H_2$)&N-U \cite{I12} ($H_2$)	&Present ($C0$)	&AIM \cite{I47} ($C0$)&N-U\cite{I12} ($C0$)\\[1ex]\hline\hline
0	&0	&4.400978574	&4.394619779	&4.39444	&11.08220218	&11.08075178	&11.08068	\\[1ex]
	&5	&4.183429818	&4.176618048	&4.17644	&11.07491243	&11.07253985	&11.07247	\\[1ex]
	&10	&3.629702678	&3.621838424	&3.62165	&11.05547579	&11.05064581	&11.05057	\\[1ex]
5	&0	&1.764237205	&1.758451567	&1.75835	&9.712515712	&9.688146187	&9.68809	\\[1ex]
	&5	&1.623366051	&1.617410615	&1.61731	&9.705448303	&9.680226284	&9.68017	\\[1ex]
	&10	&1.266759570	&1.260451640	&1.26034	&9.686604517	&9.659110919	&9.65905	\\[1ex]
7	&0	&1.082609838  &1.077636993	&1.07756	&9.191396379	&9.159164003	&9.15911	\\[1ex]
	&5	&0.966873845	&0.961814782	&0.96174	&9.184417149	&9.151359661	&9.15131	\\[1ex]
	&10	&0.675048455	&0.669844065	&0.66976	&9.165808494	&9.130552425	&9.13050	\\[1ex]
\hline\hline
\end{tabular}\label{tab2}}
\vspace*{-1pt}}
\end{table}
\begin{table}[!t]
{\scriptsize
\caption{\normalsize Ebergy spectra of Deng-Fan diatomic molecular potential for $H_2 \left(X^1\Sigma^+_g\right)$, $CO \left(X^1\Sigma^+\right)$, $HF \left(X^1\Sigma^+\right)$, $O_2 \left(X^3 \Sigma^+_g\right)$ and $O_2 ^+\left(X^2 \Pi_g\right)$ molecules for various $n$ and rotational $\ell$ quantum numbers. } \vspace*{10pt}{
\begin{tabular}{ccccccc}\hline\hline
{}&{}&{}&{}&{}&{}&{}\\[-1.0ex]
$n$&$\ell$&$H_2 \left(X^1\Sigma^+_g\right)$&$CO \left(X^1\Sigma^+\right)$&$HF \left(X^1\Sigma^+\right)$&$O_2 \left(X^3 \Sigma^+_g\right)$&$O_2 ^+\left(X^2 \Pi_g\right)$\\[2.5ex]\hline\hline
0	&0	&0.365141571630	&0.14496907267	&0.29664475468	&0.10204385069	&0.12341159004	\\[1ex]
1	&0	&0.998213655071	&0.43017899550	&0.84879137870	&0.30157505570	&0.36506351446	\\[1ex]
	&1	&1.011736491223	&0.43071963141	&0.85396890484	&0.30203765756	&0.36559505801	\\[1ex]
2	&0	&1.582601950199	&0.71191080079	&1.37474141491	&0.49742226304	&0.60257089273	\\[1ex]
	&1	&1.595035333506	&0.71244756155	&1.37972360389	&0.49788160541	&0.60309875940	\\[1ex]
	&2	&1.619836697120	&0.71352107116	&1.38968398285	&0.49880028219	&0.60415448370	\\[1ex]
3	&0	&2.107302268265	&0.98963235238	&1.86975698501	&0.68912972889	&0.83540994827	\\[1ex]
	&1	&2.118708590334	&0.99016524991	&1.87454776989	&0.68958581969	&0.83593414705	\\[1ex]
	&2	&2.141459973433	&0.99123103314	&1.88412543561	&0.69049799334	&0.83698253554	\\[1ex]
	&3	&2.175434452445	&0.99282967815	&1.89848217862	&0.69186623393	&0.83855509594	\\[1ex]
4	&0	&2.575691305740	&1.26335597760	&2.33445776235	&0.87670753899	&1.06359208920	\\[1ex]
	&1	&2.586127919535	&1.26388502379	&2.33906095729	&0.87716038610	&1.06411262902	\\[1ex]
	&2	&2.606943719809	&1.26494310432	&2.34826353550	&0.87806607249	&1.06515369967	\\[1ex]
	&3	&2.638024369517	&1.26653019550	&2.36205787817	&0.87942458223	&1.06671528329	\\[1ex]
	&4	&2.679199648264	&1.26864626172	&2.38043256798	&0.88123589159	&1.06879735299	\\[1ex]
5	&0	&2.990875171736	&1.53309394464	&2.76944475234	&1.06016573964	&1.28712867897	\\[1ex]
	&1	&3.000394872082	&1.53361915128	&2.77386405693	&1.06061535100	&1.28764556873	\\[1ex]
	&2	&3.019380399287	&1.53466955280	&2.78269894439	&1.06151456584	&1.28867933936	\\[1ex]
	&3	&3.047724487910	&1.53624512555	&2.79594197541	&1.06286336840	&1.29022997301	\\[1ex]
	&4	&3.085267435478	&1.53834583407	&2.81358200193	&1.06466173494	&1.29229744290	\\[1ex]
	&5	&3.131798515096	&1.54097163107	&2.83560417964	&1.06690963399	&1.29488171325	\\[1ex]
6	&0	&3.355714996849	&1.79885846293	&3.17530096333	&1.23951433788	&1.50603103656	\\[1ex]
	&1	&3.364366448329	&1.79937984182	&3.17953996677	&1.23996072135	&1.50654428521	\\[1ex]
	&2	&3.381618777060	&1.80042258783	&3.18801433936	&1.24085348039	&1.50757077359	\\[1ex]
	&3	&3.407371283026	&1.80198667737	&3.20071681651	&1.24219259927	&1.50911048386	\\[1ex]
	&4	&3.441474029835	&1.80407207520	&3.21763651189	&1.24397805443	&1.51116338944	\\[1ex]
	&5	&3.483729160505	&1.80667873414	&3.23875892977	&1.24620981439	&1.51372945463	\\[1ex]
	&6	&3.533892617999	&1.80980659543	&3.26406598158	&1.24888783984	&1.51680863506	\\[1ex]
7	&0	&3.672849758377	&2.06066168353	&3.55259204944	&1.41476330171	&1.72031043670	\\[1ex]
	&1	&3.680677877082	&2.06117924630	&3.55665423437	&1.41520646508	&1.72082005302	\\[1ex]
	&2	&3.696286607610	&2.06221436015	&3.56477505504	&1.41609278397	&1.72183927685	\\[1ex]
	&3	&3.719581354261	&2.06376700163	&3.57694741679	&1.41742224263	&1.72336809040	\\[1ex]
	&4	&3.750421261686	&2.06583713552	&3.59316068801	&1.41919481768	&1.72540646704	\\[1ex]
	&5	&3.788620441740	&2.06842471492	&3.61340071207	&1.42141047769	&1.72795437139	\\[1ex]
	&6	&3.833949577906	&2.07152968119	&3.63764982344	&1.42406918344	&1.73101175897	\\[1ex]
	&7	&3.886137881655	&2.07515196399	&3.66588686789	&1.42717088787	&1.73457857662	\\[1ex]
\hline\hline
\end{tabular}\label{tab3}}
\vspace*{-1pt}}
\end{table}

\end{document}